\documentclass{ws-p8-50x6-00}                                                  
                                                                               
\newcommand{\ud}{\mathrm{d}}

\begin{document}                                                               
                                                                               
\title{QUASICLASSICAL CALCULATIONS FOR WIGNER FUNCTIONS VIA      
MULTIRESOLUTION}                                                         
\author{A. Fedorova, M. Zeitlin}                                               
\address {IPME, RAS, St.~Petersburg,                                           
V.O. Bolshoj pr., 61, 199178, Russia\\                                         
e-mail: zeitlin@math.ipme.ru\\                                                 
http://www.ipme.ru/zeitlin.html;                                               
http://www.ipme.nw.ru/zeitlin.html}                                            
\maketitle                          

\abstracts{ 
We present the application of variational-wavelet analysis 
to numerical/analytical calculations of Wigner functions in (nonlinear)
quasiclassical beam dynamics problems. (Naive) deformation quantization
and multiresolution representations are the key points.
}

\section{INTRODUCTION}

In this paper we consider some starting points in the applications of a new 
numerical-analytical technique which is based on local nonlinear harmonic analysis
(wavelet analysis, generalized coherent states analysis) to the quantum/quasiclassical
(nonlinear) beam/accelerator physics calculations. The reason for this treatment is that
recently a number of problems appeared in which one needs take into account quantum properties
of particles/beams.Our starting point is the general point of view of deformation quantization approach at least on
naive Moyal/Weyl/Wigner level (part 2). The main point is that the algebras of quantum observables are the deformations
of commutative algebras of classical observables (functions) [1].So, if we have the Poisson manifold $M$ (symplectic 
manifolds, Lie coalgebras, etc) as a model for classical dynamics then for quantum calculations we need to find
an associative (but non-commutative) star product $*$ on the space of formal power series in $\hbar$ with
coefficients in the space of smooth functions on $M$ such that

\begin{eqnarray}
f * g =fg+\hbar\{f,g\}+\sum_{n\ge 2}\hbar^n B_n(f,g), 
\end{eqnarray}
where
$\{f,g\}$
is the Poisson brackets, $B_n$ are bidifferential operators 
$C^\infty(X)\otimes C^\infty(X)\rightarrow C^\infty(X) $.
There is also an infinite-dimensional gauge group on the set of star-products 
\begin{eqnarray}
f \mapsto f+\sum_{n\ge 2}\hbar^n D_n(f),
\end{eqnarray}
where
$D_n$ are differential operators. Kontsevich gave the solution to this deformation problem in terms of formal
power series via sum over graphs[1]. He also proved that for every Poisson manifold M there is a canonically
defined gauge equivalence class of star-products on M. Also there is the nonperturbative corrections to power 
series representation for $*$  [1]. In naive calculations we may use simple formal rules:

\begin{eqnarray}
* &\equiv&\exp \Big(\frac{i\hbar}{2}(\overleftarrow\partial_x\overrightarrow\partial_p-
   \overleftarrow\partial_p\overrightarrow\partial_x)\Big)\\
f(x,p)*g(x,p)&=&f(x,p-\frac{i\hbar}{2}\overrightarrow\partial_x)\cdot
g(x,p+\frac{i\hbar}{2}\overleftarrow\partial_x)\\
&=&f(x+\frac{i\hbar}{2}\overrightarrow\partial_p,p-\frac{i\hbar}{2}\overrightarrow\partial_x)
g(x,p)
\end{eqnarray}

In this paper we consider calculations of Wigner functions (WF) as the solution
of Wigner equations [2] (part 3):

\begin{eqnarray}
i\hbar\frac{\partial}{\partial t}W(x,p,t)=H * f(x,p,t)-f(x,p,t) * H
\end{eqnarray}
and especially stationary Wigner equations:

\begin{eqnarray}
H * W = W * H = Ef
\end{eqnarray}

Our approach is based on extension of our variational-wavelet approach [3]-[14].
Wavelet analysis is some set of mathematical methods, which gives us the possibility to
work with well-localized bases (Fig.~1) in functional spaces and gives maximum sparse
forms for the general type of operators (differential, integral, pseudodifferential) in such bases.
These bases are natural generalization of standard coherent, squeezed, thermal squeezed states [2],
which correspond to quadratical systems (pure linear dynamics) with Gaussian Wigner functions.
So, we try to calculate quantum corrections to classical dynamics described by polynomial nonlinear 
Hamiltonians such as orbital motion in storage rings, orbital dynamics in general multipolar
fields etc. from papers [3]-[13].
The common point for classical/quantum calculations is that
any solution
which comes from full multiresolution expansion in all space/time (or phase space)
scales represents expansion into a slow part
and fast oscillating parts (part 4).
So, we may move
from the coarse scales of resolution to the 
finest one for obtaining more detailed information about our dynamical classical/quantum process.
In this way we give contribution to our full solution
from each scale of resolution. 
The same is correct for the contribution to power spectral density
(energy spectrum): we can take into account contributions from each
level/scale of resolution.
Because affine
group of translations and dilations (or more general group, which acts on the space of solutions) 
is inside the approach
(in wavelet case), this
method resembles the action of a microscope. We have contribution to
final result from each scale of resolution from the whole underlying 
infinite scale of spaces. 
In part 5 we consider numerical modelling  of
Wigner functions 
which explicitly demonstrates quantum interference of
generalized coherent states.

\section{Quasiclassical Evolution}                                            
                                                                              
Let us consider classical and quantum dynamics in phase space               
$\Omega=R^{2m}$ with coordinates $(x,\xi)$ and generated by                   
Hamiltonian ${\cal H}(x,\xi)\in C^\infty(\Omega;R)$.                     
If $\Phi^{\cal H}_t:\Omega\longrightarrow\Omega$ is (classical) flow then
time evolution of any bounded classical observable or                      
symbol $b(x,\xi)\in C^\infty(\Omega,R)$ is given by $b_t(x,\xi)=                
b(\Phi^{\cal H}_t(x,\xi))$. Let $H=Op^W({\cal H})$ and $B=Op^W(b)$ are
the self-adjoint operators or quantum observables in $L^2(R^n)$,                
representing the Weyl quantization of the symbols ${\cal H}, b$ [1]   
\begin{eqnarray}                                                       
(Bu)(x)=\frac{1}{(2\pi\hbar)^n}\int_{R^{2n}}b\left(\frac{x+y}{2},\xi\right)
\cdot                                                                      
e^{i<(x-y),\xi>/\hbar}u(y)\ud y\ud\xi,                             
\end{eqnarray}                                           
where $u\in S(R^n)$ and $B_t=e^{iHt/\hbar}Be^{-iHt/\hbar}$ be the    
Heisenberg observable or quantum evolution of the observable $B$       
under unitary group generated by $H$. $B_t$ solves the Heisenberg equation of
motion                                                             
$\dot{B}_t=({i}/{\hbar})[H,B_t].$                                        
Let $b_t(x,\xi;\hbar)$ is a symbol of $B_t$ then we have      
 the following equation for it                                        
\begin{equation}                                                             
\dot{b}_t=\{ {\cal H}, b_t\}_M,                                    
\end{equation}                                    
with the initial condition $b_0(x,\xi,\hbar)=b(x,\xi)$.             
Here $\{f,g\}_M(x,\xi)$ is the Moyal brackets of the observables
$f,g\in C^\infty(R^{2n})$, $\{f,g\}_M(x,\xi)=f\sharp g-g\sharp f$,
where $f\sharp g$ is the symbol of the operator product and is presented
by the composition of the symbols $f,g$                           
\begin{eqnarray}
&&(f\sharp g)(x,\xi)=\frac{1}{(2\pi\hbar)^{n/2}}\int_{R^{4n}}
e^{-i<r,\rho>/\hbar+i<\omega,\tau>/\hbar}\\
&& \cdot f(x+\omega,\rho+\xi)\cdot
g(x+r,\tau+\xi)\ud\rho \ud\tau \ud r\ud\omega \nonumber
\end{eqnarray}
For our problems it is useful that $\{f,g\}_M$ admits the formal
expansion in powers of $\hbar$:
\begin{eqnarray}
\{f,g\}_M(x,\xi)\sim \{f,g\}+2^{-j}\cdot
\sum_{|\alpha+\beta|=j\geq 1}(-1)^{|\beta|}\cdot
(\partial^\alpha_\xi fD^\beta_x g)\cdot(\partial^\beta_\xi
 gD^\alpha_x f),
\end{eqnarray}
 where $\alpha=(\alpha_1,\dots,\alpha_n)$ is
a multi-index, $|\alpha|=\alpha_1+\dots+\alpha_n$,
$D_x=-i\hbar\partial_x$.
So, evolution (1) for symbol $b_t(x,\xi;\hbar)$ is
\begin{eqnarray}
\dot{b}_t=\{{\cal H},b_t\}+\frac{1}{2^j}
\sum_{|\alpha +\beta|=j\geq 1}(-1)^{|\beta|}
\cdot
\hbar^j
(\partial^\alpha_\xi{\cal H}D_x^\beta b_t)\cdot
(\partial^\beta_\xi b_t D_x^\alpha{\cal H}).
\end{eqnarray}
At $\hbar=0$ this equation transforms to classical Liouville equation.
Equation (12) plays the key role in many quantum (semiclassical) problems. We consider its 
particular case--Wigner equation--in the next section.

\section{Wigner Equations}
According to Weyl transform quantum state (wave function or density operator) corresponds
to Wigner function, which is analog of classical phase-space distribution [2].
We consider the following form of differential equations for time-dependent WF

\begin{eqnarray}
\partial_tW(p,q,t)=\frac{2}{\hbar}\sin\Big[\frac{\hbar}{2}
(\partial^H_q\partial^W_p-\partial^H_p\partial^W_q)\Big]\cdot H(p,q)W(p,q,t)
\end{eqnarray}
Let
\begin{eqnarray}
\hat{\rho}=|\Psi_\epsilon><\Psi_\epsilon|\
\end{eqnarray}
be the density operator or projection operator corresponding to the energy eigenstate
$|\Psi_\epsilon>$ with energy eigenvalue $\epsilon$. Then time-independent Schroedinger equation
corresponding to Hamiltonian
\begin{eqnarray}
\hat{H}(\hat{p},\hat{q})=\frac{\hat{p}^2}{2m}+U(\hat{q})
\end{eqnarray}
where $U(\hat{q}$ is arbitrary polynomial function (related beam dynamics models
considered in [3]-[13]) on $\hat{q}$ is [2]:
\begin{eqnarray}
\hat{H}\hat{\rho}=\epsilon\hat{\rho}
\end{eqnarray}
After Weyl-Wigner mapping we arrive at the following equation on WF in c-numbers:
\begin{eqnarray}
H\big(p+\frac{\hbar}{2i}\frac{\partial}{\partial q},q-\frac{\hbar}{2i}\frac{\partial}{\partial p}\Big)W(p,q)
  =\epsilon W(p,q)
\end{eqnarray}
or
\begin{eqnarray*}
\Big( \frac{p^2}{2m}+\frac{\hbar}{2i}\frac{p}{m}\frac{\partial}{\partial q}-
 \frac{\hbar^2}{8m}\frac{\partial^2}{\partial q^2}\Big)W(p,q)+
 U\Big(q-\frac{\hbar}{2i}\frac{\partial}{\partial p}\Big)W(p,q)=\epsilon W(p,q)
\end{eqnarray*}
 After expanding the potential $U$ into the Taylor series we have two real partial differential equations
\begin{eqnarray}
\Big(-\frac{p}{m}\frac{\partial}{\partial q}+\sum_{m=0}^\infty
\frac{1}{(2m+1!)}\Big(\frac{i\hbar}{2}\Big)^{2m}
\frac{\ud^{2m+1}U}{\ud q^{2m+1}}\frac{\partial^{2m+1}}{\partial p^{2m+1}}\Big)W(p,q)=0
\end{eqnarray}

\begin{eqnarray}
&&\Big(\frac{p^2}{2m}+U(q)-\frac{\hbar^2}{8m}\frac{\partial^2}{\partial q^2}+
 \sum_{n=1}^\infty\frac{1}{(2n)!}\Big(\frac{i\hbar}{2}\Big)^{2n}
 \frac{\ud^{2n}U}{\ud q^{2n}}\frac{\partial^{2n}}{\partial p^{2n}}\Big) W(p,q)= \nonumber\\
&&\epsilon W(p,q)
\end{eqnarray}
In the next section we consider  variation-wavelet approach for the solution of these 
equations for the case of arbitrary polynomial $U(q)$, which corresponds to a finite number 
of terms in equations (18), (19) up to any order of $\hbar$.

\section{Variational Multiscale Representation}
Let L be arbitrary (non)linear differential operator with matrix dimension $d$, 
which acts on some set of functions
$\Psi\equiv\Psi(x,y)=\Big(\Psi^1(x,y),...,\Psi^d(x,y)\Big), \quad x,y\in\Omega\subset\Re^2$
from $L^2(\Omega)$:
\begin{equation}
L\Psi\equiv L(Q,x,y)\Psi(x,y)=0,
\end{equation}
where
\begin{equation}
Q\equiv Q_{d_1,d_2,d_3,d_4}(x,y,\partial /\partial x,\partial /\partial y)=
\sum_{i,j,k,\ell=1}^{d_1,d_2,d_3,d_4}a_{ijk\ell}x^iy^j
\Big(\frac{\partial}{\partial x}\Big)^k\Big(\frac{\partial}{\partial y}\Big)^\ell
\end{equation}
Let us consider now the N mode approximation for solution as the following ansatz (in the same way
we may consider different ansatzes):
\begin{equation}
\Psi^N(x,y)=\sum^N_{r,s=1}a_{r,s}\Psi_r(x)\Phi_s(y)
\end{equation}
We shall determine coefficients of expansion from the following Galerkin conditions
(different related variational approaches are considered in [3]-[13]):
\begin{equation}
\ell^N_{k\ell}\equiv\int(L\Psi^N)\Psi_k(x)\Phi_\ell(y)\ud x\ud y=0
\end{equation}
So, we have exactly $dN^2$ algebraical equations for  $dN^2$ unknowns $a_{rs}$.

But in the case of equations for WF (18), (19) we have overdetermined system of equations:
$2N^2$ equations for $N^2$ unknowns $a_{rs}$ (in this case $d=1$).
In this paper we consider non-standard method for resolving this problem,
which is based on biorthogonal wavelet expansion. So, instead of expansion (22) we consider
the following one:
\begin{equation}
\Psi^N(x,y)=\sum^N_{r,s=1}a_{r,s}\Psi_r(x)\Psi_s(y)+\sum^N_{i,j=1}\widetilde{a}_{ij}
\widetilde{\Psi}_i(x)\widetilde{\Phi}_j(y),
\end{equation}
where $\widetilde{\Psi}_i(x)\widetilde{\Phi}_j(y)$ are the bases dual to initial ones.
Because wavelet functions are the generalization of coherent states we consider 
an expansion on this overcomplete set of bases wavelet functions as a generalization of 
standard coherent states expansion.

So, variational/Galerkin approach reduced the initial problem (20) to the problem of solution 
of functional equations at the first stage and some algebraical problems at the second
stage. We'll consider now the multiresolution expansion as the second main part of our 
construction. 
Because affine                                             
group of translation and dilations is inside the approach, this             
method resembles the action of a microscope. We have contribution to        
final result from each scale of resolution from the whole
infinite scale of increasing closed subspaces $V_j$:
$
...V_{-2}\subset V_{-1}\subset V_0\subset V_{1}\subset V_{2}\subset ...
$.
The solution is parametrized by solutions of two reduced algebraical
problems, one is linear or nonlinear (23)
(depends on the structure of operator L)  and the second one are some linear
problems related to computation of coefficients of algebraic equations (23).
These coefficients can be found  by the method of Connection
Coefficients (CC)[15] or related method [16].
We use compactly supported wavelet basis functions for expansions (22), (24).
We may consider different types of wavelets including general wavelet packets (section 5 below).
These coefficients depend on the wavelet-Galerkin integrals. In general
we need to find ($d_i\ge 0$)
\begin{eqnarray}
\Lambda^{d_1 d_2 ...d_n}_{\ell_1 \ell_2 ...\ell_n}=
 \int\limits_{-\infty}^{\infty}\prod\varphi^{d_i}_{\ell_i}(x)dx
\end{eqnarray}
According to CC method [15] we use the next construction for quadratic case. When $N$  in
scaling equation is a finite even positive integer the function
$\varphi(x)$  has compact support contained in $[0,N-1]$.
For a fixed triple $(d_1,d_2,d_3)$ only some  $\Lambda_{\ell
 m}^{d_1 d_2 d_3}$ are nonzero: $2-N\leq \ell\leq N-2,\quad
2-N\leq m\leq N-2,\quad |\ell-m|\leq N-2$. There are
$M=3N^2-9N+7$ such pairs $(\ell,m)$. Let $\Lambda^{d_1 d_2 d_3}$
be an M-vector, whose components are numbers $\Lambda^{d_1 d_2
d_3}_{\ell m}$. Then we have the following reduced algebraical system
: $\Lambda$
satisfy the system of equations $(d=d_1+d_2+d_3)$
\begin{eqnarray}
A\Lambda^{d_1 d_2 d_3}=2^{1-d}\Lambda^{d_1 d_2 d_3},
\quad
A_{\ell,m;q,r}=\sum\limits_p a_p a_{q-2\ell+p}a_{r-2m+p}
\end{eqnarray}
By moment equations we have created a system of $M+d+1$
equations in $M$ unknowns. It has rank $M$ and we can obtain
unique solution by combination of LU decomposition and QR
algorithm.
For nonquadratic case we have analogously additional linear problems for
objects (25).
Solving these linear problems we obtain the coefficients of reduced main linear/nonlinear
algebraical system (23) and after its solution we obtain the coefficients of wavelet
expansion (22), (24). 
As a result we obtained the explicit  solution  of our
problem in the base of compactly supported wavelets (22).

 Also
in our case we need to consider
the extension of this approach to the case of
any type of variable coefficients (periodic, regular or singular).
We can produce such approach if we add in our construction additional
refinement equation, which encoded all information about variable
coefficients [16].
So, we need to compute only additional
integrals of
the form
\begin{equation}\label{eq:var1}
\int_Db_{ij}(t)(\varphi_1)^{d_1}(2^m t-k_1)(\varphi_2)^{d_2}
(2^m t-k_2)\ud x,
\end{equation}
where  $b_{ij}(t)$ are arbitrary functions of time and trial
functions $\varphi_1,\varphi_2$ satisfy the refinement equations:
\begin{equation}
\varphi_i(t)=\sum_{k\in{\bf Z}}a_{ik}\varphi_i(2t-k)
\end{equation}
If we consider all computations in the class of compactly supported wavelets
then only a finite number of coefficients do not vanish. To approximate
the non-constant coefficients, we need choose a different refinable function
$\varphi_3$ along with some local approximation scheme
\begin{equation}
(B_\ell f)(x):=\sum_{\alpha\in{\bf Z}}F_{\ell,k}(f)\varphi_3(2^\ell t-k),
\end{equation}
where $F_{\ell,k}$ are suitable functionals supported in a small neighborhood
of $2^{-\ell}k$ and then replace $b_{ij}$ in (27) by
$B_\ell b_{ij}(t)$.  To guarantee
sufficient accuracy of the resulting approximation to (27)
it is important to have the flexibility of choosing $\varphi_3$ different
from $\varphi_1, \varphi_2$. 
 So, if we take
$\varphi_4=\chi_D$, where $\chi_D$ is characteristic function of $D$, 
which is again a refinable function, then the problem of
computation of (\ref{eq:var1}) is reduced to the problem of calculation of
integral
\begin{eqnarray}
&&H(k_1,k_2,k_3,k_4)=H(k)=
\int_{{\bf R}^s}\varphi_4(2^j t-k_1)\cdot \nonumber\\
&&\varphi_3(2^\ell t-k_2)
\varphi_1^{d_1}(2^r t-k_3)
\varphi_2^{d_2}(2^st-k_4)\ud x
\end{eqnarray}
The key point is that these integrals also satisfy some sort of algebraical
equation [16]:
\begin{equation}
2^{-|\mu|}H(k)=\sum_{\ell\in{\bf Z}}b_{2k-\ell}H(\ell),\qquad \mu=d_1+d_2.
\end{equation}
This equation can be interpreted as the problem of computing an eigenvector.
Thus, the problem of extension of our approach to the case of
variable coefficients is reduced to the same standard
algebraical problem as in
case of constant coefficients. So, the general scheme is the same one and we
have only one more additional
linear algebraic problem. After solution of these linear problems  we can again 
compute coefficients of wavelet expansions (22), (24).

Now we concentrate on the last additional problem  which comes from 
overdeterminity of equations (18), (19), which demands to consider expansion (24)
instead of expansion (22). It leads to equal number of equations and 
unknowns in reduced algebraical system of equations (23). For this reason we consider
biorthogonal wavelet analysis.
We started with two hierarchical sequences of approximations spaces [16]:
$
\dots V_{-2}\subset V_{-1}\subset V_{0}\subset V_{1}\subset V_{2}\dots,$
$\dots \widetilde{V}_{-2}\subset\widetilde{V}_{-1}\subset
\widetilde{V}_{0}\subset\widetilde{V}_{1}\subset\widetilde{V}_{2}\dots, 
$
and as usually,
$W_0$ is complement to $V_0$ in $V_1$, but now not necessarily orthogonal
complement.
New orthogonality conditions have now the following form:
$
\widetilde {W}_{0}\perp V_0,\quad  W_{0}\perp\widetilde{V}_{0},\quad
V_j\perp\widetilde{W}_j, \quad \widetilde{V}_j\perp W_j,
$
translates of $\psi$ $\mathrm{span}$ $ W_0$,
translates of $\tilde\psi \quad \mathrm{span} \quad\widetilde{W}_0$.
Biorthogonality conditions are
$
<\psi_{jk},\tilde{\psi}_{j'k'}>=
\int^\infty_{-\infty}\psi_{jk}(x)\tilde\psi_{j'k'}(x)\ud x=
\delta_{kk'}\delta_{jj'},
$
 where
$\psi_{jk}(x)=2^{j/2}\psi(2^jx-k)$.
Functions $\varphi(x), \tilde\varphi(x-k)$ form  dual pair:
$
<\varphi(x-k),\tilde\varphi(x-\ell)>=\delta_{kl},\quad
 <\varphi(x-k),\tilde\psi(x-\ell)>=0\quad  \mbox{for}\quad \forall k,
\ \forall\ell.
$
Functions $\varphi, \tilde\varphi$ generate a multiresolution analysis.
$\varphi(x-k)$, $\psi(x-k)$ are synthesis functions,
$\tilde\varphi(x-\ell)$, $\tilde\psi(x-\ell)$ are analysis functions.
Synthesis functions are biorthogonal to analysis functions. Scaling spaces
are orthogonal to dual wavelet spaces.
Two multiresolutions are intertwining
$
V_j+W_j=V_{j+1}, \quad \widetilde V_j+ \widetilde W_j = \widetilde V_{j+1}
$.
These are direct sums but not orthogonal sums.
So, our representation for solution has now the form
$
f(t)=\sum_{j,k}\tilde b_{jk}\psi_{jk}(t),
$
where synthesis wavelets are used to synthesize the function. But
$\tilde b_{jk}$ come from inner products with analysis wavelets.
Biorthogonal point of view is more flexible and stable under the action of large class
of operators while orthogonal (one scale for multiresolution) is fragile,
all computations are much more simple and we accelerate the rate of convergence of our 
expansions (24). By analogous anzatzes and approaches we may construct also the 
multiscale/multiresolution representations for solution of time dependent Wigner equation (13) [14].

\begin{figure}[htb]
\epsfxsize=10pc
\figurebox{10pc}{10pc}{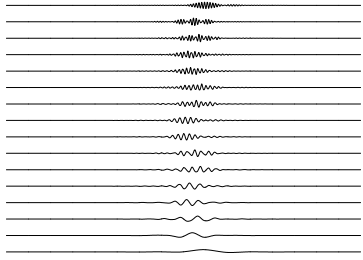}
\caption{Localized contributions to beam motion.}
\end{figure}

\section{Numerical Modelling}
So, our constructions give us the following N-mode representation for solution of Wigner equations
(18)-(19):

\begin{equation}
W^N(p,q)=\sum^N_{r,s=1}a_{rs}\Psi_r(p)\Phi_s(q)
\end{equation}
where $\Psi_r(p)$, $\Phi_s(q)$ may be represented by some family of (nonlinear)
eigenmodes with the corresponding multiresolution/multiscale representation in the
high-localized wavelet bases (Fig.~1): 

\begin{eqnarray}
\Psi_k(p)&=&\Psi^{M_1}_{k, slow}(p)+\sum_{i\ge M_1}\Psi^i_k(\omega_i^1p),
\quad \omega^1_i \sim 2^i\\
\Phi_k(q)&=&\Phi^{M_2}_{k, slow}(q)+\sum_{j\ge M_2}\Phi^j_k(\omega_j^2q),
\quad \omega^2_j \sim 2^j
\end{eqnarray}
Our (nonlinear) eigenmodes are more realistic for the modelling of 
nonlinear classical/quantum dynamical process  than the corresponding linear gaussian-like
coherent states. Here we mention only the best convergence properties of expansions 
based on wavelet packets, which  realize the so called minimal Shannon entropy property (Fig.~1).
On Fig.~2 we present numerical modelling [17] of Wigner function for a simple model of beam motion,
which explicitly demonstrates quantum interference property. On Fig.~3 we present 
the multiscale/multiresolution representation (32)-(34) for solution of Wigner equation.

\newpage
\begin{figure}[htb]
\epsfxsize=20pc
\figurebox{20pc}{20pc}{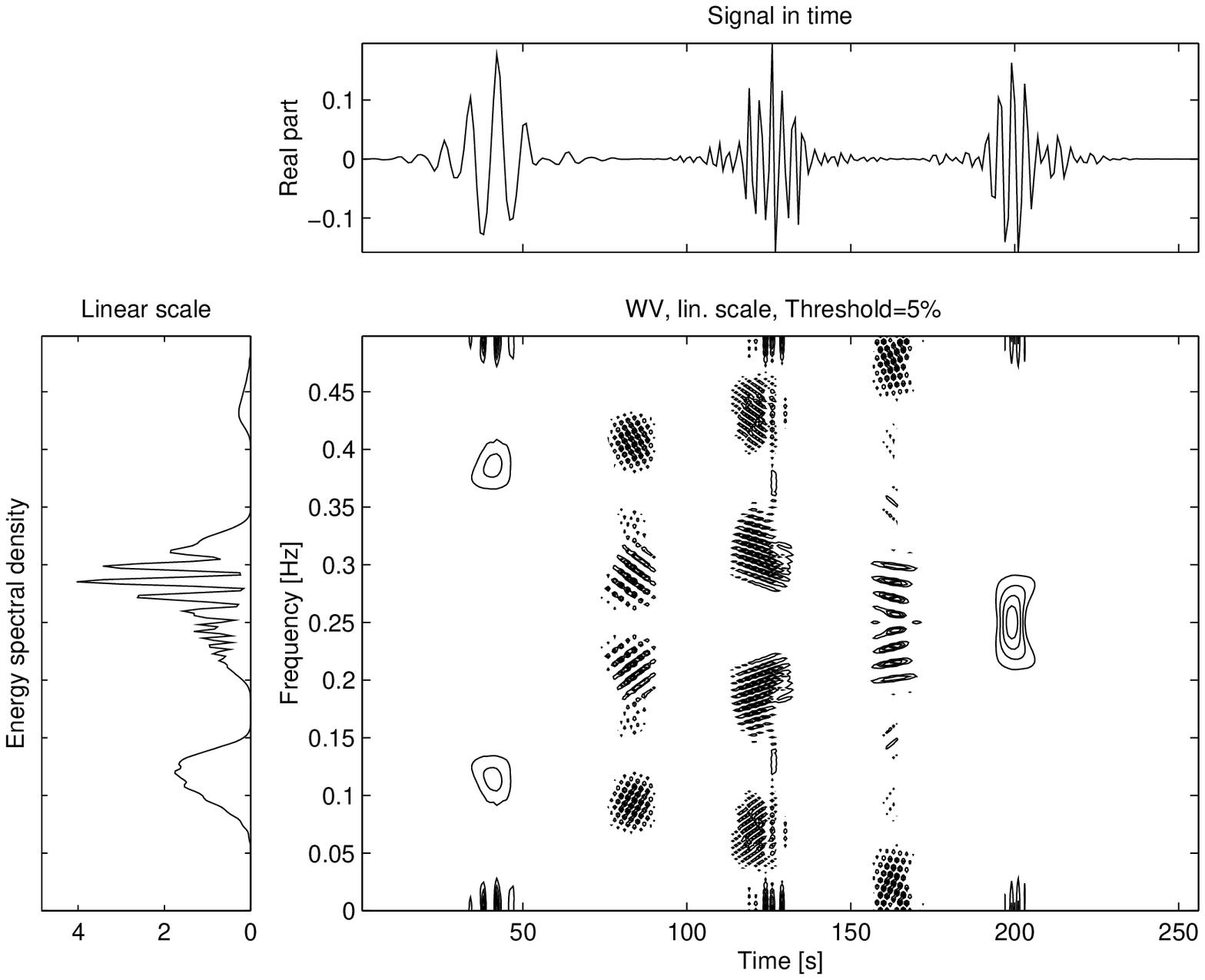}
\end{figure}

\begin{figure}[htb]
\epsfxsize=20pc
\figurebox{25pc}{25pc}{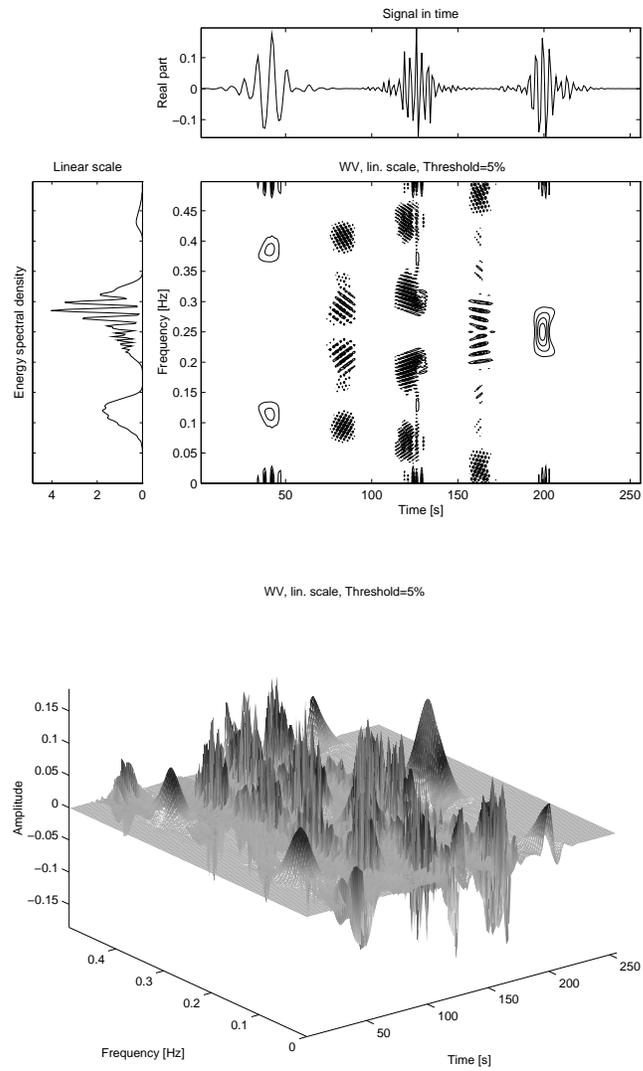}
\caption{Wigner function for 3 wavelet packets.}
\end{figure}

\newpage

\begin{figure}[htb]
\epsfxsize=20pc
\figurebox{25pc}{25pc}{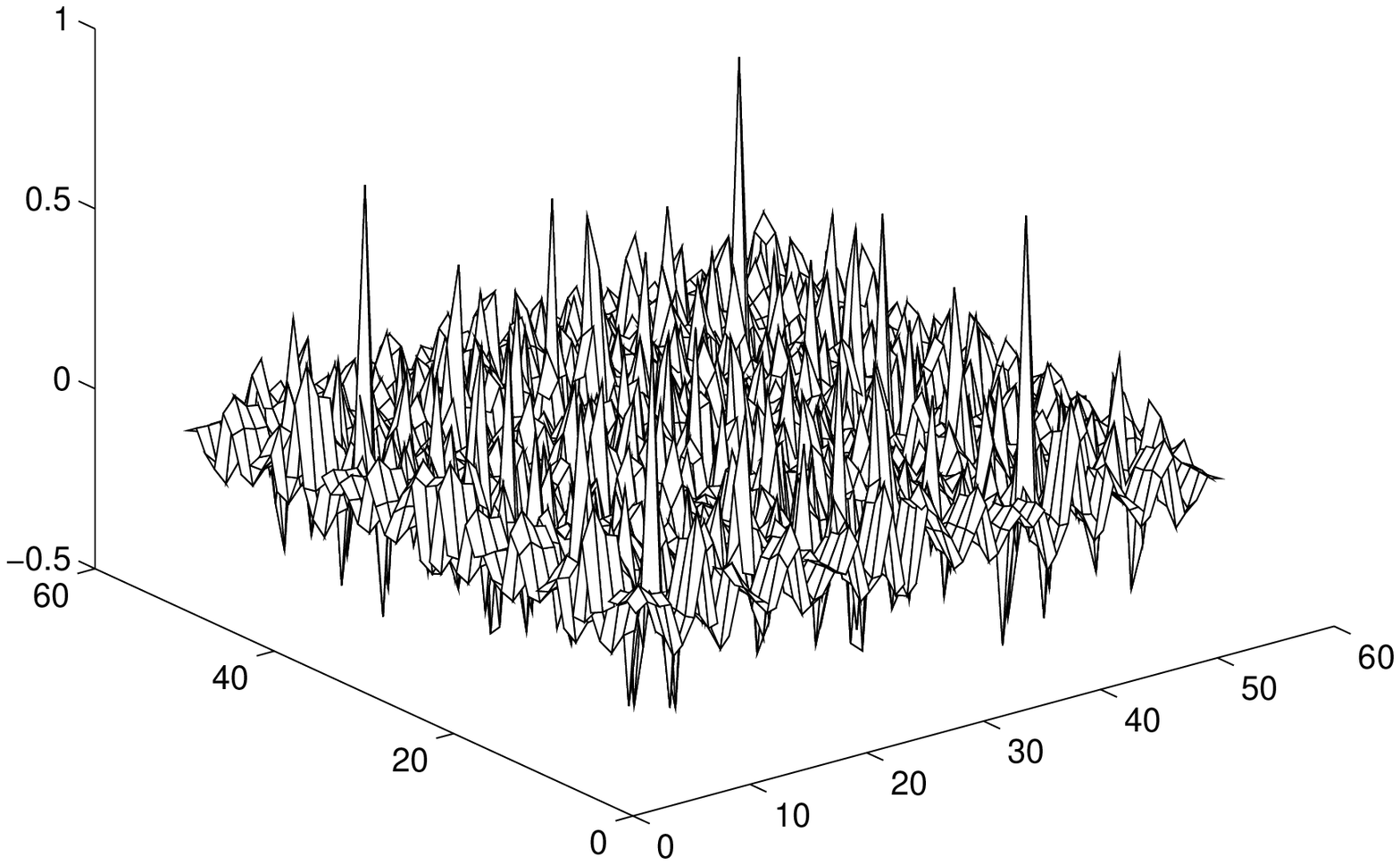}
\end{figure}

\begin{figure}[htb]
\epsfxsize=20pc
\figurebox{25pc}{25pc}{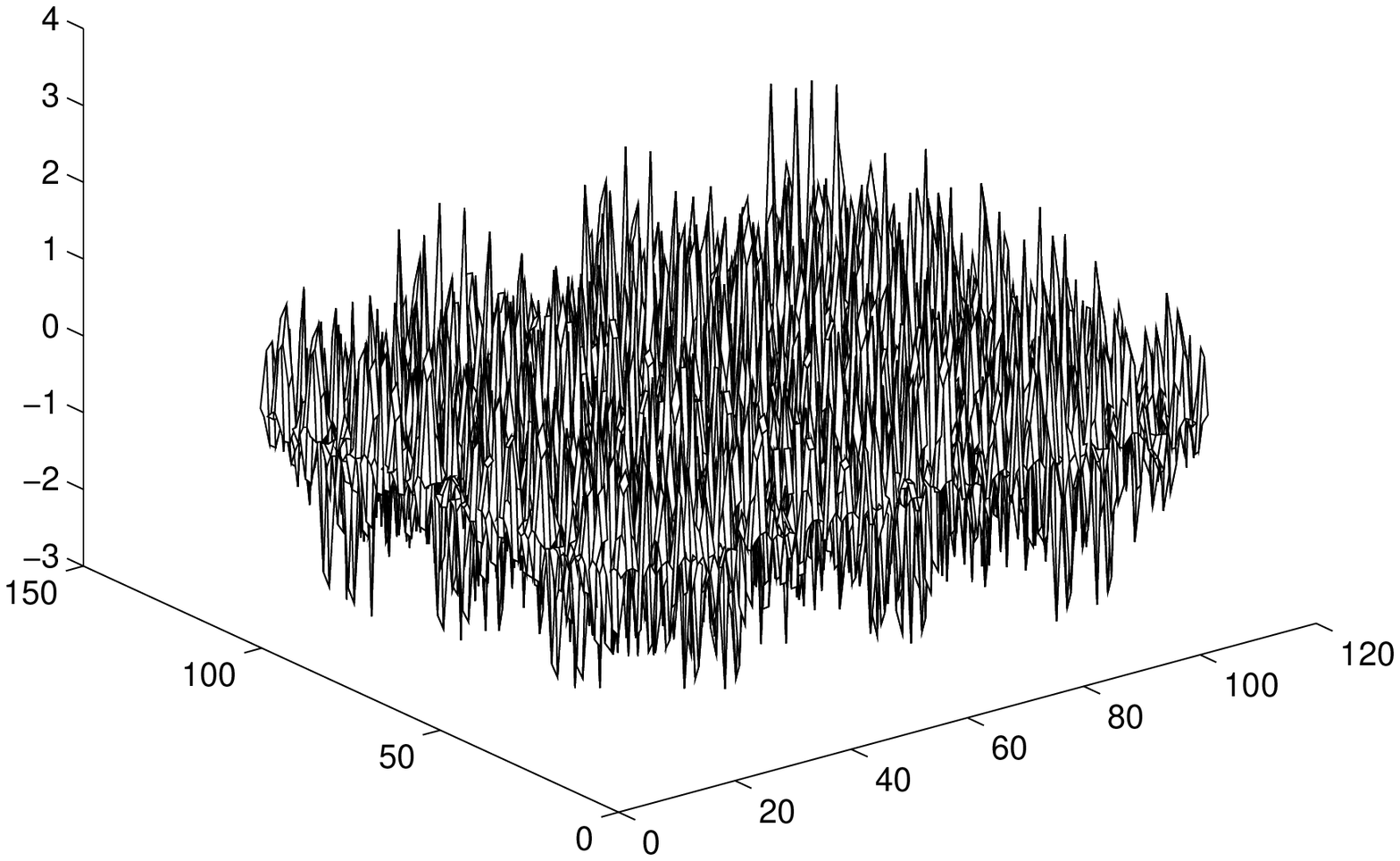}
\caption{Multiresolution/multiscale representations for Wigner functions.}
\end{figure}

We would like to thank Professor
Pisin Chen, Dr. Stefania Petracca and her team for nice hospitality, help and support
during Capri ICFA Workshop.


\newpage                                                                       
\begin{thebibliography}{17}                                                    
                                                                               
\bibitem{1}                                                                    
D. Sternheimer, Los Alamos preprint: math.QA/9809056, 
M. Kontsevich, q-alg/9709040,
V. Periwal, hep-th/0006001.                                                       
                                                                                      
\bibitem{2}                                                                         
T. Curtright, T. Uematsu, C. Zachos, hep-th/0011137, M.Huq, e.a., 
{\it Phys. Rev.}, {\bf A 57}, 3188 (1998).
          
                                                                           
                                                                                  
\bibitem{3}                                                                           
A.N. Fedorova and M.G. Zeitlin,                                                       
 {\it Math. and Comp. in Simulation}, {\bf 46}, 527 (1998).                           
                                                                                      
                                                                                      
\bibitem{4}                                                                           
A.N. Fedorova and M.G. Zeitlin,                                           
'Wavelet Approach to Mechanical Problems. Symplectic Group,                           
Symplectic Topology and Symplectic Scales',                                           
{\it New Applications of Nonlinear and Chaotic Dynamics in Mechanics}, 31, 101        
(Kluwer,  1998).                                                                      
                                                                                   
\bibitem{5}                                                                           
A.N. Fedorova and M.G. Zeitlin,                                                       
{\bf CP405}, 87 (American Institute of Physics, 1997).                                
Los Alamos preprint, phy\-sics/9710035.

\bibitem{6}
A.N. Fedorova, M.G. Zeitlin and Z.~Parsa,
Proc. PAC97
{\bf 2}, 1502, 1505, 1508 (IEEE, 1998).
\bibitem{7}
A.N. Fedorova, M.G. Zeitlin and Z.~Parsa,
Proc. EPAC98, 930, 933 (Institute of Physics, 1998).

\bibitem{8}
A.N. Fedorova, M.G. Zeitlin and Z.~Parsa,
{\bf CP468}, 48 (American Institute of Physics, 1999).
Los Alamos preprint, physics/990262.

\bibitem{9}
A.N. Fedorova, M.G. Zeitlin and Z.~Parsa,
{\bf CP468}, 69 (American Institute of Physics, 1999).
Los Alamos preprint, physics/990263.

 \bibitem{10}
A.N. Fedorova and M.G. Zeitlin,
Proc. PAC99,
1614, 1617, 1620, 2900, 2903,
2906, 2909, 2912 (IEEE/APS, New York, 1999).\\
Los Alamos preprints:
physics/9904039, 9904040, 9904041, 9904042, 9904043,
9904045, 9904046, 9904047.

\bibitem{11}
A.N. Fedorova and M.G. Zeitlin,
Proc. UCLA ICFA Workshop, in press,
Los Alamos preprint: physics/0003095.

\bibitem{12}
A.N. Fedorova and M.G. Zeitlin,  Proc. EPAC00, 415, 872,  1101, 1190, 1339, 2325.\\
Los Alamos preprints: physics/0008045, 0008046, 0008047, 0008048, 0008049, 0008050

\bibitem{13}
A.N. Fedorova, M.G. Zeitlin, Proc.  LINAC00, 2 papers in press,
Los Alamos preprints: physics/0008043, 0008200

\bibitem{14}
A.N. Fedorova, M.G. Zeitlin, this Volume and in press.

\bibitem{15}
A. Latto, e.a. Aware Technical Report AD910708 (1991).

\bibitem{16}
W. Dahmen, C. Micchelli, {\it SIAM J. Numer. Anal.}, {\bf 30}, 507 (1993).

\bibitem{17}
F. Auger, e.a., Time-frequency Toolbox, CNRS/Rice Univ. (1996).


\end{thebibliography}
 \end{document}